# Common Radio Resource Management Policy for Multimedia Traffic in Beyond 3G Heterogeneous Wireless Systems

M.Carmen Lucas-Estañ(1), Javier Gozalvez(1), Joaquin Sanchez-Soriano(2)
(1)Ubiquitous Wireless Communications Research (Uwicore) Laboratory (2)Operations Research Center
University Miguel Hernandez
Elche, Spain
j.gozalvez@umh.es

***Abstract*—Beyond 3G wireless systems will be composed of a variety of Radio Access Technologies (RATs) with different, but also complementary, performance and technical characteristics. To exploit such diversity while guaranteeing the interoperability and efficient management of the different RATs, common radio resource management (CRRM) techniques need to be defined. This work proposes and evaluates a CRRM policy that simultaneously assigns to each user an adequate combination of RAT and number of radio resources within such RAT to guarantee its QoS requirements. The proposed CRRM technique is based on linear objective functions and programming tools.**

## I. INTRODUCTION

There is a wide consensus in the research community that Beyond 3G wireless systems will consist of heterogeneous mobile and wireless Radio Access Technologies (RATs) physically coexisting, but with distinct performance and technical characteristics. Ensuring their interoperability would allow an efficient use of their common radio resources to guarantee the user required Quality of Service (QoS) levels and the maximum systems' revenue. To achieve these objectives, a key feature of heterogeneous Beyond 3G wireless systems will be the definition of adequate Common Radio Resource Management (CRRM) policies. CRRM techniques are aimed at deciding the RAT over which to convey a user transmission (RAT selection), and the number of resources from the selected RAT (intra-RAT RRM) that the transmission would require to satisfy the QoS requirements. Several studies separately tackle the RAT selection and the intra-RAT RRM dilemmas. For example, [1] and [2] propose intra-RAT RRM policies based on utility functions and economic laws. While the authors in [1] suggest applying auctions principles to maximize the operator's benefit, [2] proposes the use of bankruptcy-based resource allocation policies to guarantee user fairness and avoid channel access stagnation under heterogeneous traffic environments. In terms of RAT selection techniques, [3] proposes a general framework for their definition, and some specific examples based on pre-established service/RAT assignments. Additional RAT selection principles based on signal strength and instantaneous loads are proposed and evaluated in [4].

This work has been supported by the Ministry of Education and Science (Spain) and FEDER funds under the projects TEC2005-08211-C02-02 and MTM2005-09184-C02-02.

Initial proposals to jointly address the RAT selection and intra-RAT RRM dilemmas have been recently proposed. For example, [5] proposes a CRRM algorithm based on neural networks and fuzzy logic that simultaneously determines the most appropriate RAT and bit rate allocation, considering, among other factors, the user QoS constraints. While [5] determines the necessary bit rate at the assigned RAT, it does not tackle the problem of intra-RAT radio resources assignment. In this context, this work proposes and evaluates a CRRM policy that simultaneously assigns to each user an adequate combination of RAT and number of radio resources within such RAT to guarantee the user/service QoS requirements. The proposed algorithm is based on linear programming optimization techniques, which have shown significant benefits in other areas, and have been recently applied for the first time to CRRM in [6] to address the RAT selection dilemma. The work reported in [6] proposes a joint call admission control scheme for minimizing call blocking probability in heterogeneous wireless networks. In this context, [6] used linear programming optimization techniques to determine the optimal splitting of arriving calls among available RATs. The objectives of the work reported in [6] significantly differ to those of this paper, where a CRRM policy jointly addressing the RAT selection and intra-RAT RRM dilemmas is proposed.

## II. UTILITY FUNCTIONS

The operation of the proposed RAT and resource assignment policy is based on service-dependent utility functions. These utility functions define the user-perceived QoS for a varying number of assigned radio resources per available RAT. As demonstrated in [2], properly chosen utility functions are capable to adequately reflect the user-perceived QoS levels. In addition, these utility functions can allow for a dynamic and efficient policy-dependent resource assignment that achieves its QoS objectives independently of the system load. Given the length restrictions, this paper will not describe in detail the process followed to establish the utility functions, but will rather present its main characteristics.

This work considers an heterogeneous multimedia traffic scenario with email, web and real-time H.263 video (with different bit rates) users. The utility values (Fig. 1) have been established considering minimum, medium and maximum QoS requirements (Table I). While web and email utility values refer to achievable throughputs, the video utility figure corresponds to the percentage of video frames transmitted before the next video frame is to be transmitted. This video utility representation has been chosen to define utility functions independent of the H.263 video bit rates. For all services, utility values equal to zero are assigned if the minimum QoS requirement cannot be achieved. Although the use of utility values to estimate the user satisfaction is a subjective process, it is important to note that the authors demonstrated in [2] that the utility functions reported in Fig. 1 adequately reflect the user needs, validating the procedure employed to define them.

Once the utility functions are established, it is then necessary to define the relation between radio resources and utility values. Since this work focuses on heterogeneous wireless systems (GPRS, EDGE and HSDPA), such relation would need to consider the selected RAT and the number of radio resources within such RAT that are been assigned. GPRS and EDGE systems have been implemented with a single carrier (i.e. eight timeslots) each. For HSDPA, this work considers the transmission modes defined in the CQI (Channel Quality Indicator) mapping table for User Equipment category 10, which does not consider transmission modes employing 6, 9, 11, 13 and 14 codes, and defines a total of 30 different transmission modes [7]; each transmission mode corresponds to a combination of codes, modulation and coding schemes.

To relate the utility functions defined in Fig. 1 to the various possible RAT and resources combination, a transmission rate is selected per radio resource in each RAT. For real-time H.263 video users, an additional step is necessary. A cumulative distribution function (cdf) of the throughput needed to transmit each video frame before the next video frame is to be transmitted is derived following the H.263 video model proposed in [8]. With this cdf, the percentage of video frames reported in Fig. 1 can be related to the corresponding necessary throughputs for the various video bit rates considered in this work (16, 64, 128, 256 and 512kbps).

It is important to note that all emulated RATs implement Adaptive Modulation and Coding (AMC), which can then result in varying, and difficult to predict, transmission rates for the same number of radio resources. In this scenario, it has been chosen to consider transmission modes providing a balance between data rates and error correction. In particular, average data rates of 13.4 kbps (corresponding to the coding scheme CS2) and 22.4kbps (corresponding to the modulation and coding scheme MCS5) have been selected for a timeslot in GPRS and EDGE, respectively. In HSDPA, various transmission modes can be defined per assigned code. To achieve the sought data rate/error correction balance, the selected transmission rate per assigned HSDPA code corresponds to that achieved by the 'intermediate' transmission mode out of all possible modes for a given number of HSDPA codes. Such transmission modes would correspond to CQI value 3 for 1 HSDPA code, and CQI 8 for 2 HSDPA codes in the example shown in Table II.

Once an average data rate is established per radio resource for all possible RATs, it is then possible to represent the utility values that can be achieved for each RAT and number of resources being assigned. Table III shows an example for real-time 64 kbps H.263 video users with the RAT/resources combination shown in increasing throughput order. The resources/RAT combination is denoted as $xY$, corresponding to $x$ radio resources (timeslots or codes) from RAT $Y$ (GPRS is represented as $G$, EDGE as $E$, and HSDPA as $H$). It is interesting to note that certain RAT/resources combinations cannot achieve utility values greater than zero, and that from a certain RAT/resources combination, the utility value does not increase despite raising transmission throughputs.

TABLE I. USER QOS LEVELS

| | Min. QoS | Mean QoS | Max. QoS |
|---|---|---|---|
| WWW | 32kbps | 64kbps | 128kbps |
| Email | 16kbps | 32kbps | 64kbps |
| Established utility values | 0.95/4 | 0.95/2 | 0.95 |
| H.263 video | 75% | 95% | 100% |
| Established utility values | 0.95/4 | 0.95/2 | 1 |

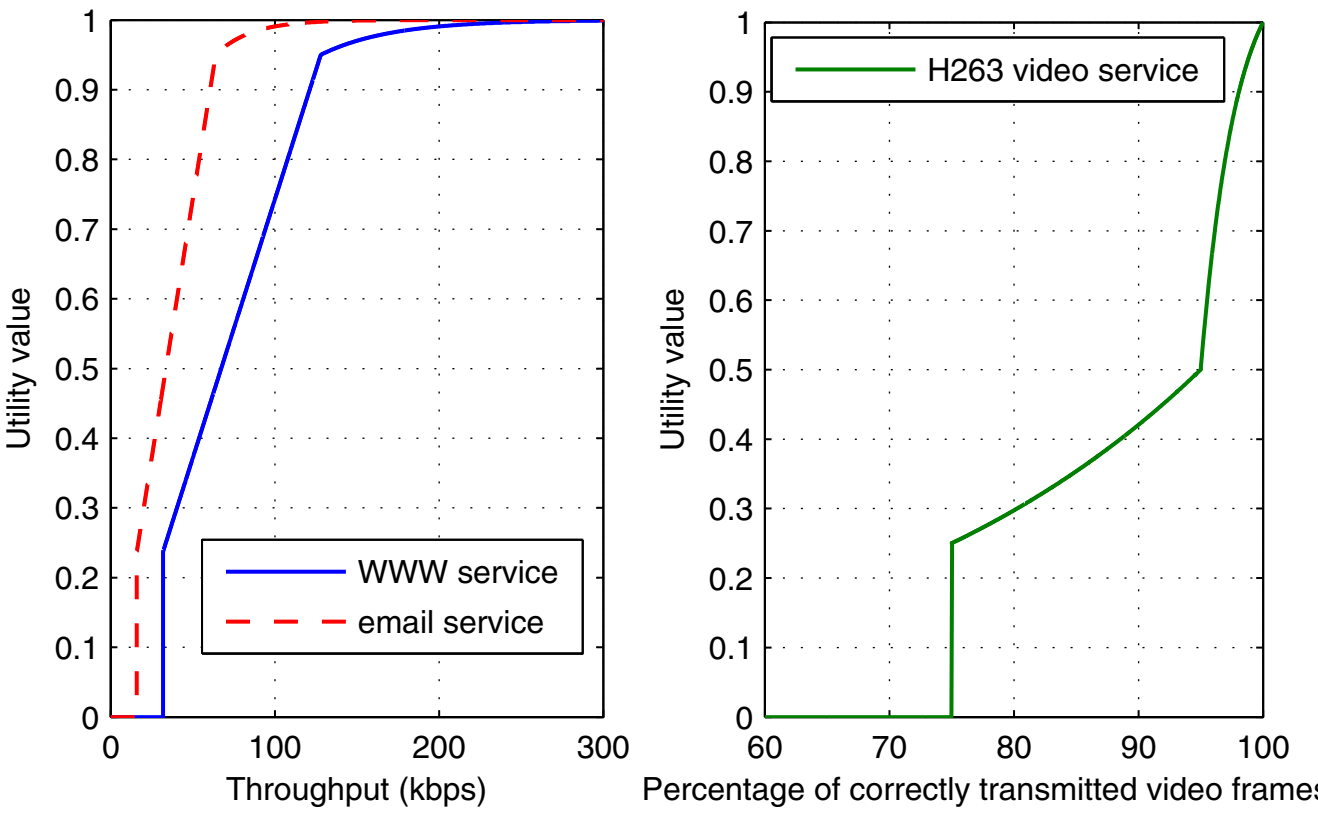

Figure 1. Utility functions per traffic service.

TABLE II. EXTRACT OF CQI MAPPING TABLE FOR UE CATEGORY 10

| CQI value | Data Rate (kbps) | Codes | CQI value | Data Rate (kbps) | Codes |
|---|---|---|---|---|---|
| 1 | 68.5 | 1 | 6 | 230.5 | 1 |
| 2 | 86.5 | 1 | 7 | 325 | 2 |
| 3 | 116.5 | 1 | 8 | 396 | 2 |
| 4 | 158.5 | 1 | 9 | 465.5 | 2 |
| 5 | 188.5 | 1 | | | |

TABLE III. 64KBPS VIDEO UTILITY VALUES

| Res./ RAT | Data rate (kbps) | Utility value | Res./ RAT | Data rate (kbps) | Utility value | Res./ RAT | Data rate (kbps) | Utility value |
|---|---|---|---|---|---|---|---|---|
| 1G | 13.4 | 0.0000 | 4E | 89.6 | 0.2983 | 3H | 741 | 1.0000 |
| 1E | 22.4 | 0.0000 | 7G | 93.8 | 0.3127 | 4H | 1139.5 | 1.0000 |
| 2G | 26.8 | 0.0000 | 8G | 107.2 | 0.3532 | 5H | 2332 | 1.0000 |
| 3G | 40.2 | 0.0000 | 5E | 112 | 0.3654 | 7H | 4859.5 | 1.0000 |
| 2E | 44.8 | 0.0000 | 1H | 116.5 | 0.3775 | 8H | 5709 | 1.0000 |
| 4G | 53.6 | 0.0000 | 6E | 134.4 | 0.4350 | 10H | 7205.5 | 1.0000 |
| 5G | 67 | 0.0000 | 7E | 156.8 | 0.9338 | 12H | 8618.5 | 1.0000 |
| 3E | 67.2 | 0.0000 | 8E | 179.2 | 0.9819 | 15H | 11685 | 1.0000 |
| 6G | 80.4 | 0.0000 | 2H | 396 | 1.0000 | | | |

## III. Common Radio Resource Assignment Proposal

This work is aimed at defining a CRRM policy that maximizes the channel's efficiency while guaranteeing the adequate user-perceived QoS levels based on the established operator's service policy. As it has been previously explained, the proposed policy jointly addresses the RAT selection and the decision on the number of radio resources within the selected RAT that need to be assigned to each user. However, it is important to note that this work does not yet consider the possibility that one user is simultaneously assigned resources from various RATs. The proposed policy is based on the previously established utility functions and linear programming optimization tools. While it is evaluated under a fairness user policy with service priorisation when radio resources are limited under high system loads, the proposed CRRM mechanism could be easily extended to other operator's strategies.

### *A. CRRM objective function*

As previously explained, each service type has different resource requirements to achieve similar QoS levels. Such requirements have been defined by means of the utility functions described in Section II, where it was shown that similar user satisfaction levels (represented here by means of utility values) can be achieved through various resources/RAT combinations. The proposed CRRM policy exploits such flexibility to decide on the optimum resources/RAT assignments following a user fairness approach. In fact, the proposed technique aims at providing similar, and highest possible, user satisfaction levels for all service types. Only when the number of available radio resources is lower than the demand will the implemented policy priorise certain traffic classes. In this context, the proposed CRRM objective function can be denoted as follows:

$$\max \prod_j u_j, \; j \in [0, N-1] \tag{1}$$

which is equivalent to:

$$\max \; \ln \prod_j u_j, \quad j \in [0, N-1] \tag{2}$$

where $u_j$ represents the utility value assigned to user $j$ in a combined RAT/resources distribution round, and $N$ corresponds to the total system user load. Under equal service and operative constraints, (1) or (2) is achieved when utility values are equally distributed among users. The established CRRM objective function then can be expressed as:

$$\ln \prod_j u_j = \sum_j \ln u_j \tag{3}$$

$$u_j = \sum_r \sum_s u_j^r\left(s^r\right) \cdot y_j^{r,s} \tag{4}$$

with $u_j$ defined in (4). In (4), $u^r_j(s^r)$ represents the utility value obtained by user $j$ when assigned $s$ radio resources (codes or timeslots) of RAT $r$ ($r$ is equal to 0, 1 or 2 for GPRS, EDGE and HSDPA respectively), and $s \in [1,c_r]$ with $c_r$ corresponding to the maximum number of radio resources available at each RAT. $y_j^{r,s}$ is a binary variable equal to one if user $j$ is assigned $s$ radio resources of RAT $r$, and equal to 0 if not. The proposed CRRM policy focuses then on deciding for each user which $y_j^{r,s}$ variable is equal to one, considering that only $y_j^{r,s}$ variables achieving an utility value greater than zero are allowed. Given that only one $y_j^{r,s}$ variable can be equal to one for each user, the following expression applies:

$$\sum_j \ln u_j = \sum_j \ln \sum_r \sum_s u_j^r\left(s^r\right) \cdot y_j^{r,s} = \sum_j \sum_r \sum_s \ln\left(u_j^r\left(s^r\right) \cdot y_j^{r,s}\right) \tag{5}$$

If we assume that all users must have a variable $y_j^{r,s}$ equal to one, (5) becomes:

$$\sum_j \sum_r \sum_s \ln\left(u_j^r\left(s^r\right) \cdot y_j^{r,s}\right) = \sum_j \sum_r \sum_s \ln\left(u_j^r\left(s^r\right)\right) \cdot y_j^{r,s} \tag{6}$$

Our CRRM objective function can then be expressed as:

$$\max \sum_j \sum_r \sum_s \ln\left(u_j^r\left(s^r\right)\right) \cdot y_j^{r,s} \tag{7}$$

### *B. CRRM service constraints*

Given that (7) is a linear equation, it can then be resolved using linear programming tools. However, before describing the linear programming resolution, it is important to complete the problem statement with some service-specific constraints. To obtain a linear objective function, it has been assumed that each user will obtain a $y_j^{r,s}$ variable equal to one. However, this condition might not be feasible under high system loads where the number of radio resources is lower than the user demands. In this case, the number of users requesting resources will be reduced so that:

$$\sum_r \sum_s y_j^{r,s} = 1, \forall j \tag{8}$$

$$\sum_j \sum_{s^r} s^r \cdot y_j^{r,s} \le c_r, \forall r \tag{9}$$

Whenever a user requests resources for a new transmission or a given transmission ends, the developed CRRM policy is applied. In this case, only real-time video active users that were assigned resources in the previous CRRM distribution round maintain those corresponding to the minimum QoS level and compete for additional ones, which can be expressed as:

$$\sum_{r}\sum_{s^r} E_{s^r} \cdot y_j^{r,s} \geq E_{\min}, \forall j_{h263} \quad (10)$$

where $E_{s^r}$ represents the RAT/resources combination index in utility tables (e.g. Table III), corresponding to the assignment of $s$ radio resources for RAT $r$. Similarly, $E_{min}$ represents the index of the RAT/resources combination achieving the minimum QoS level.

In case the available resources do not allow achieving equal utility values for all users, users are served based on the following service priorisation: real-time H.263 video (higher priority), web, and email. Within the real-time video service class, users with higher video bit rates are served first. If the lowest priority user ($m$) is a video user that obtained radio resources in the previous CRRM distribution round, the condition established in (10) comes first and the video user would be assigned the $s_{min}$ radio resources from RAT $r_{min}$ necessary to achieve the minimum QoS level defined by $E_{min}$. When such level is achieved, the lowest priority user will not be assigned additional resources until the highest priority user ($k$) surpasses its utility value ($u_m^{r_{\min}}(s_{m_{\min}})$). This condition can be summarized in (11):

$$\sum_{r_a}\sum_{s_a} u_m^{r_{\min}}\left(s_{m_{\min}}\right) \cdot y_k^{r,s} + \sum_{r_b}\sum_{s_b} u_k^{r}\left(s^r\right) \cdot y_k^{r,s} \geq \sum_{r}\sum_{s} u_m^{r}\left(s^r\right) \cdot y_m^{r,s} \quad (11)$$

This condition is only applied when the priority of user $k$ is higher than that of user $m$, where ($r_a$,$s_a$) represent the RAT/resources assignments that verify $u_m^{r_{\min}}\left(s_{m_{\min}}\right) - u_k^{r_a}\left(s_a\right) > 0$ and ($r_b$,$s_b$) the assignments that verify $u_m^{r_{\min}}\left(s_{m_{\min}}\right) - u_k^{r_b}\left(s_b\right) \leq 0$.

It has been previously explained that to obtain a linear objective function, it is necessary that all users participating in the CRRM resources distribution round obtain a $y_j^{r,s}$ variable equal to one. If this is not possible due to resources shortage, it was mentioned that the number of users requesting resources will be reduced to satisfy all imposed constraints. Following (11), it can be established that if present users cannot even obtain their minimum QoS demand, the linear objective function does not have a solution, and users with the lowest priority are eliminated from the CRRM distribution round until such solution can be achieved.

### C. CRRM linear programming resolution

Linear programming tools have been shown to represent an attractive solution in optimization problems looking at identifying the most suitable solution following a previously established objective function. In this context, this work proposes its use to address the CRRM problem in heterogeneous wireless systems, where policies are needed to determine which RAT, and with how many radio resources, is assigned to each user.

The previous section has mathematically expressed the problem statement with all its system and service constraints (III.B.), and has derived a linear objective function to solve, with a binary integer unknown variable $y_j^{r,s}$ . In operations research, this type of problems is referred to as Binary Integer Programming (BIP) [9], and can be solved using different approaches. One of the most popular approaches due to its performance and computational properties, and the one used in this work, is the Branch and Bound method [9]. This technique solves an ordered sequence of reduced linear programming problems until an optimum solution is achieved. Such reduced linear programming problems are obtained when the condition that the unknown variable must be an integer variable is relaxed and real variables are allowed. To solve such reduced linear programming problems, this work proposes to use the simplex method [9], which is regularly employed in problems, with a large number of variables, which require computationally efficient solutions. The simplex method is an algebraic procedure that makes use of the fact that the system and user constraints present in the problem statement reduce the range of possible solutions to a limited spatial region. The interested reader is referred to [9] for additional details on the simplex method and its operation.

## IV. CRRM PERFORMANCE

### A. Simulation conditions

The performance of the proposed CRRM technique has been evaluated in Matlab. The implemented simulator is not aimed at accurately modeling radio transmissions, but at measuring the efficiency of the resource distribution proposals and optimizing them according to each RAT operational characteristics and specific system constraints. The implemented platform emulates the distribution of GPRS, EDGE and HSDPA resources among real-time H.263 video, email and web users for various system loads. In terms of service distribution, email, web and real-time video transmissions represent each one third of the total load. Within the real-time H.263 video system load, users are equally distributed among three different bit rates selected from the emulated ones: 16, 64, 128, 256 and 512kbps. A single cell with same GPRS, EDGE and HSDPA coverage is modelled.

### B. Results

The proposed CRRM policy has been evaluated under two different multimedia traffic scenarios. In the first one (E1), web, email and real-time H.263 video transmissions at 16, 64 and 128kbps bit rates are emulated. The second evaluation scenario (E2) also considers web and email transmissions, but real-time H.263 video applications operate at 64, 256 and 512kbps bit rates. For both scenarios, cell loads of 20 and 30 users have been simulated following the traffic distributions reported in Section IV.A

Fig. 2 depicts the achieved utility values for E1 under a 20 and 30 users cell load, respectively. The figure shows the percentage of users per service class that achieved the utility values corresponding to the minimum, medium and maximum QoS levels defined in II. To understand why maximum QoS levels cannot be achieved for all service classes, it is important to note the limited available resources per RAT, and the fact that several service classes required more than one radio resource to obtain a utility value greater than zero. In this context, Fig. 2 shows that the implemented CRRM policy achieves its various objectives; for example, the fact that under resource shortage scenarios such as the ones simulated, the service priorities criteria defined in (11) is correctly applied. In this respect, it should be emphasized that the discrete nature of the utility functions preclude the possibility to achieve identical utility levels for all service types, and the traffic priorities previously defined need to be applied. It is also very important to note that the majority of services achieve their minimum QoS levels, and only when such levels are guaranteed, resources are additionally assigned to higher priority users. This does not always apply to the lowest priority services (i.e. email and web) due to the resource shortages, the service priorisation and the service continuity applied to video transmissions that were assigned resources in previous CRRM distribution rounds.

Table IV shows, per service class, the selected RAT and the assigned number of radio resources within the selected RAT for the E1 scenario under the two emulated cell loads, and for the E2 scenario under a 20 users cell load. In this table, '0RS' indicates a user has not been assigned any resource. The results depicted in Table IV show that when the cell load increases, the combined RAT/radio resources distribution is modified. In particular, the CRRM policy looks to satisfy a higher percentage of users with the highest possible QoS level, instead of satisfying a lower percentage of users with their higher QoS requirements. This approach results in that, as the load increases, the distribution better adjusts the assignments to the minimum and medium QoS requirements in order to provide the minimum QoS requirements for the higher possible percentage of users. If we look at the 64kbps video transmissions, Table IV shows that under a load of 20 users per cell in the E1 scenario, 76% of its transmissions have been assigned 2 HSDPA codes and the minimum resource assignment corresponds to 1H for 10.5% of the transmissions. According to Table III, 2H and 1H provided utility values of 1 and 0.38 for 64kbps video transmissions, which shows that all transmissions guarantee their minimum QoS requirements but when possible higher QoS levels are assigned. As the load increases to 30 users per cell, the lowest 64kbps resource assignment is 4E (just above the minimum QoS requirements) and the percentage of users with higher utility values decreases, highlighting the CRRM resource distribution adaptation with the traffic load. A similar trend to that observed when increasing the load is also experienced when increasing the service QoS requirements. In fact, a similar resources distribution is obtained when passing from the E1 scenario with a 20 users load to the E2 scenario with the same system load. The higher QoS requirements of video services under the E2 scenario force a new resources distribution that, for example, reduces the resources assignment (and its corresponding utility values) for the 64kbps video, web and email users; these services also increase their probability of not being assigned any radio resource due to the resource shortage caused by the higher minimum QoS requirements of higher bit rate video services.

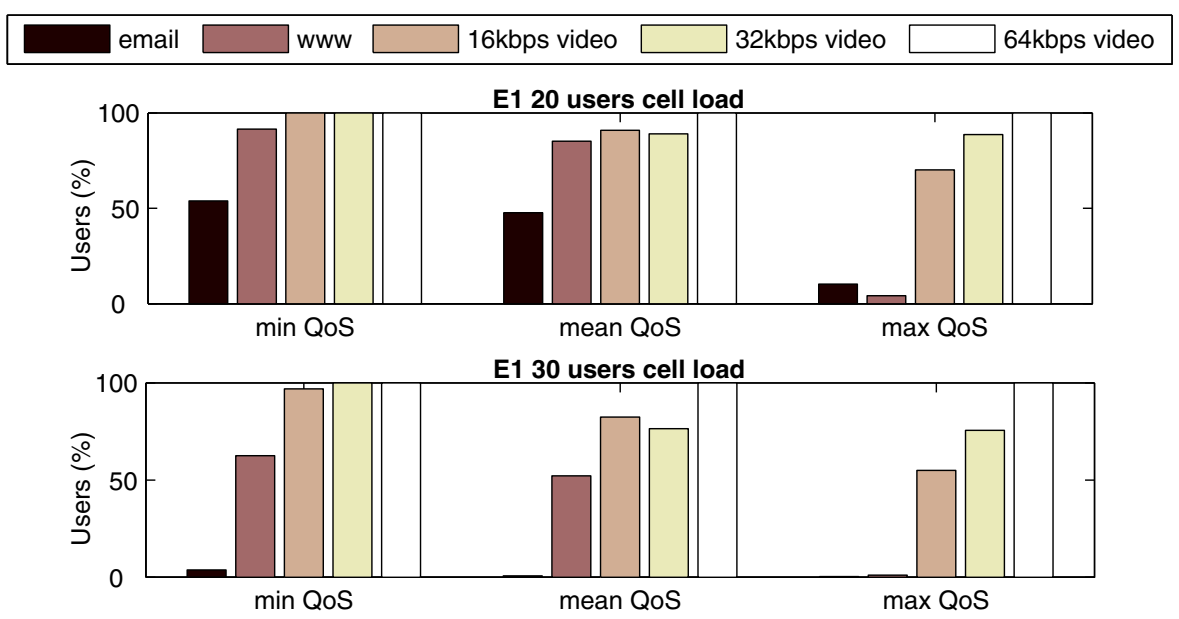

Figure 2. Assigned utility values per service class.

TABLE IV. RADIO RESOURCE ASSIGNMENT PER SERVICE CLASS

| | E1 – 20 users | | E1 – 30 users | | | E2 – 20 users | |
|---|---|---|---|---|---|---|---|
| | **Res.** | **%** | **Res.** | **%** | | **Res.** | **%** |
| email | 0RS | 46.1 | 0RS | 96.3 | email | 0RS | 58.8 |
| | 2G | 4.9 | 1E | 2.1 | | 2G | 7.0 |
| | 3G | 6.0 | 2G | 0.9 | | 3G | 2.5 |
| | 2E | 24.4 | 3G | 0.1 | | 2E | 16.9 |
| | 4G | 6.9 | 2E | 0.2 | | 4G | 5.4 |
| | 1H | 10.3 | 1H | 0.2 | | 1H | 7.9 |
| www | 0RS | 8.5 | 0RS | 37.5 | www | 0RS | 17.4 |
| | 2E | 4.6 | 2E | 7.8 | | 3G | 2.8 |
| | 8G | 1.4 | 4G | 1.7 | | 2E | 8.4 |
| | 1H | 78.4 | 5G | 2.2 | | 4G | 1.6 |
| | 8E | 1.7 | 8G | 0.9 | | 8G | 2.6 |
| | 2H | 1.8 | 1H | 45.6 | | 1H | 64.2 |
| 16kbps video | 0RS | 0.0 | 0RS | 3.2 | 64kbps video | 0RS | 5.6 |
| | 2G | 9.1 | 2G | 10.8 | | 4E | 4.1 |
| | 3G | 20.8 | 3G | 27.4 | | 1H | 27.6 |
| | 2E | 15.0 | 2E | 30.0 | | 8E | 11.1 |
| | 1H | 43.1 | 1H | 17.1 | | 2H | 44.9 |
| 64kbps video | 0RS | 0.0 | 0RS | 0.0 | 256kbps video | 0RS | 5.6 |
| | 1H | 10.5 | 4E | 2.3 | | 2H | 9.0 |
| | 7E | 0.4 | 1H | 16.8 | | 3H | 52.0 |
| | 8E | 12.7 | 8E | 8.5 | | 4H | 6.6 |
| | 2H | 75.9 | 2H | 67.0 | | 5H | 26.6 |
| 128kbps video | 0RS | 0.0 | 0RS | 0.0 | 512kbps video | 0RS | 3.2 |
| | 2H | 52.0 | 2H | 55.8 | | 4H | 14.3 |
| | 3H | 48.0 | 3H | 44.2 | | 5H | 57.1 |
| | | | | | | 7H | 15.1 |